\documentclass[USenglish]{article}

\usepackage[utf8]{inputenc}
\usepackage[small]{dgruyter}
\usepackage{microtype}
\usepackage{subcaption}

\usepackage{tikz}

\newcommand{\leadingzero}[1]{\ifnum #1<10 0\the#1\else\the#1\fi}
\newcommand{\mytoday}{\leadingzero{\day}.\leadingzero{\month}.\the\year} 

\newcommand{\tableHeader}{\hline\noalign{\smallskip}}
\newcommand{\tableFooter}{\hline\noalign{\smallskip}}

\newcommand{\myvector}[1]{\mathbf{#1}}

\newcommand{\bulkVelocityC}{V}
\newcommand{\bulkVelocity}{\mathbf{\veloTFC}}

\newcommand{\bulkPressure}{P}

\newcommand{\dt}{{\partial_t}}

\newcommand{\Grad}{\nabla}

\newcommand{\vecLaplace}{\boldsymbol{\Delta}}

\newcommand{\R}{\mathbb{R}}

\newcommand{\I}{\myvector{I}}
\newcommand{\trace}{\operatorname{trace}}

\renewcommand{\Re}{\mathrm{Re}}

\newcommand{\insertColorbarVertical}[5]{
	\begin{minipage}{1.5cm}
		\begin{flushleft}
			\begin{tikzpicture}
				\node (colorbar) at (0,0) {\includegraphics[width=0.6cm]{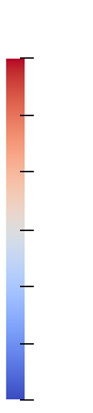}};
				\draw (0.01,1.4) node {\scriptsize #1};
				\draw (-0.15,0.965) node[anchor=west] {\scriptsize #5};
				\draw (-0.15,0.22666667) node[anchor=west] {\scriptsize #4};
				\draw (-0.15,-0.5116667) node[anchor=west] {\scriptsize #3};
				\draw (-0.15,-1.25) node[anchor=west] {\scriptsize #2};
			\end{tikzpicture}
		\end{flushleft}
	\end{minipage}
}

\newcommand{\veloTFC}{\bulkVelocityC}

\newcommand{\QTFC}{Q}
\newcommand{\QTF}{\boldsymbol{\QTFC}}

\newcommand{\pressTF}{\bulkPressure}

\newcommand{\nablaTF}{\nabla}

\newcommand{\divTF}{\nablaTF\cdot}

\newcommand{\stress}{\boldsymbol{\sigma}}

\newcommand{\activeStress}{\stress^{\textup{A}}}

\newcommand{\berisStress}{\stress^{\textup{B}}}

\newcommand{\hTFC}{H}
\newcommand{\hTF}{\boldsymbol{\hTFC}}
\newcommand{\DTFC}{D}
\newcommand{\DTF}{\boldsymbol{\DTFC}}

\newcommand{\OmegaTFC}{\Omega}
\newcommand{\OmegaTF}{\boldsymbol{\OmegaTFC}}

\begin{document}


  \author[1]{D. Wenzel}
  \author[1]{M. Nestler}
  \author[1]{S. Reuther}
  \author[1]{M. Simon}
  \author*[1,2]{A. Voigt} 
  \runningauthor{D. Wenzel et al.}
  \affil[1]{Institute of Scientific Computing, TU Dresden, 01062 Germany}
  \affil[2]{Center for Systems Biology Dresden (CSBD), Pfotenhauerstr. 108, 01307 Dresden, Germany and Cluster of Excellence - Physics of Life, TU Dresden, 01062 Dresden, Germany}
  \title{Defects in active nematics – algorithms for identification and tracking}
  \runningtitle{Defects in active nematics}
  \abstract{The growing interest in active nematics and the emerging evidence of the relevance of topological defects in biology asks for reliable data analysis tools to identify, classify and track such defects in simulation and microscopy data. We here provide such tools and demonstrate on two examples, on an active turbulent state in an active nematodynamic model and on emerging nematic order in a multi-phase field model, the possibility to compare statistical data on defect velocities with experimental results. The considered tools, which are physics based and data driven, are compared with each other.
  }
  \keywords{active liquid crystals, tensor fields, topological defects, neural networks}

\maketitle

\section{Introduction} 
\label{sec1}

Over the last years several theoretical and experimental model systems have been developed to study the collective behaviour of active matter. These are systems which extract energy from their surroundings at the single unit level and transform it into mechanical work, see \cite{Marchettietal_RMP_2013,Menzel_PR_2015,Prostetal_NP_2015,Juelicheretal_RPP_2018} for general reviews. One well-studied unit are rod-shaped particles, which include, for example, elongated bacteria and filamentous particles inside living cells. These active matter systems bear a resemblance to nematic liquid crystals, systems which are characterised by long-range orientational order. Similar nematic ordering has also been found in epithelia tissue, where fairly isotropic living cells align to each other. For both types of these natural materials current developments to gain a better understanding of the mechanisms at work are thus based on established theories for liquid crystals. We refer to \cite{Doostmohammadietal_NC_2018} for a current review on active nematics. Just like liquid crystals these active materials also show topological defects. These are regions where the nematic order is lost  in order to minimize stresses. In 2D, defects are characterized by a topological charge (in the mathematics community also often called the winding number), i.e. the angle by which the orientational direction rotates around the defect, divided by $2\pi$. This quantity is additive, conserved, and determined by the topology of the confinement. Energetically most favourable defects in nematic liquid crystals are $\pm \frac{1}{2}$ defects. 

The main difference with liquid crystals, however, lies in the ‘activity’ of the active matter, which leads to the spontaneous generation/annihilation of these topological defects. In contrast with passive systems, where defects annihilate each other, depending on the strength of the activity and the nature of the system activity might even lead to turbulence. It has been proposed that $+ \frac{1}{2}$ defects in active nematics behave as self-propelled particles and its velocity is proportional to activity \cite{Giomietal_PTRS_2014}. This relation would provide a way to measure activity in such natural materials. 

It is becoming evident that these defects also play a biological role. Defects with topological charge $+ \frac{1}{2}$ drive the dynamics and have a strong elastic dipole, while the defects with charge $- \frac{1}{2}$ are moved around passively. The motion of defects not only allows to measure activity, it also provides a good way of distinguishing extensile and contractile materials: for extensile systems $+ \frac{1}{2}$ defects move towards their ‘head’, as has been shown experimentally for microtubule (MT) bundles \cite{Sanchezetal_N_2012}, human bronchial epithelial cells \cite{Blanch-Mercaderetal_PRL_2018} and Madine–Darby canine kidney (MDCK) cells \cite{Sawetal_N_2017}, whereas for contractile systems they move towards their ‘tail’, which has been observed in experiments on mouse fibroblast cells \cite{Duclosetal_NP_2017}. In \cite{Sawetal_N_2017} it was found that near $+ \frac{1}{2}$ defects the rate of apoptosis of MDCK epithelial cells is higher, due to the presence of isotropic compressive stresses. In contrast, the $- \frac{1}{2}$ defects are characterized by tensile stresses and do not trigger apoptosis. In \cite{Kawaguchietal_N_2017} collective dynamics of cultured murine neural progenitor cells (NPCs) are studied. At high density the cells were capable of forming an aligned pattern. Rapid cell accumulation at $+ \frac{1}{2}$ defects and escape from $- \frac{1}{2}$ defects has been identified. 

For all these interpretations a robust identification and tracking of defects in microscopy as well as simulation data is essential. We will here focus on these issues and describe physics based as well as neural network based approaches to identify topological defects. The paper is structured as follows: In Section \ref{sec2} we introduce two essential mathematical models for active nematics. One, a hydrodynamic nematic liquid crystal model with added activity, and the other a collective model of active cells, where their interaction leads to nematic ordering. In Section \ref{sec3} we focus on $\pm \frac{1}{2}$ topological defects and describe the used algorithms to locate them, to identify the topological charge and to track them over time. In Section \ref{sec4} results are shown for various examples and similarities and differences between the two models are highlighted. In section \ref{sec5} conclusions are drawn. 

\section{Mathematical description} 
\label{sec2}

We consider two theoretical model systems which lead to active nematics. Both consider a continuous description but they describe different levels of detail. The first is a coarse-grained model for rod-shaped particles based on a Landau-de Gennes $Q$-tensor theory, e.g. suitable to model MT bundles. The second describes each unit by a phase field variable and is appropriate to model epithelia cells. The interaction between the units leads to shape deformations and cell elongation, from which again a Q-tensor can be computed. The numerical solution of these systems of partial differential equations is based on finite elements and implemented in AMDiS \cite{Veyetal_CVS_2007,Witkowskietal_ACM_2015}. 

\subsection{Active nematodynamics}
\label{ssec21}

\begin{figure}[h]
\center
\begin{minipage}{0.36\textwidth}
    \includegraphics[width=\textwidth]{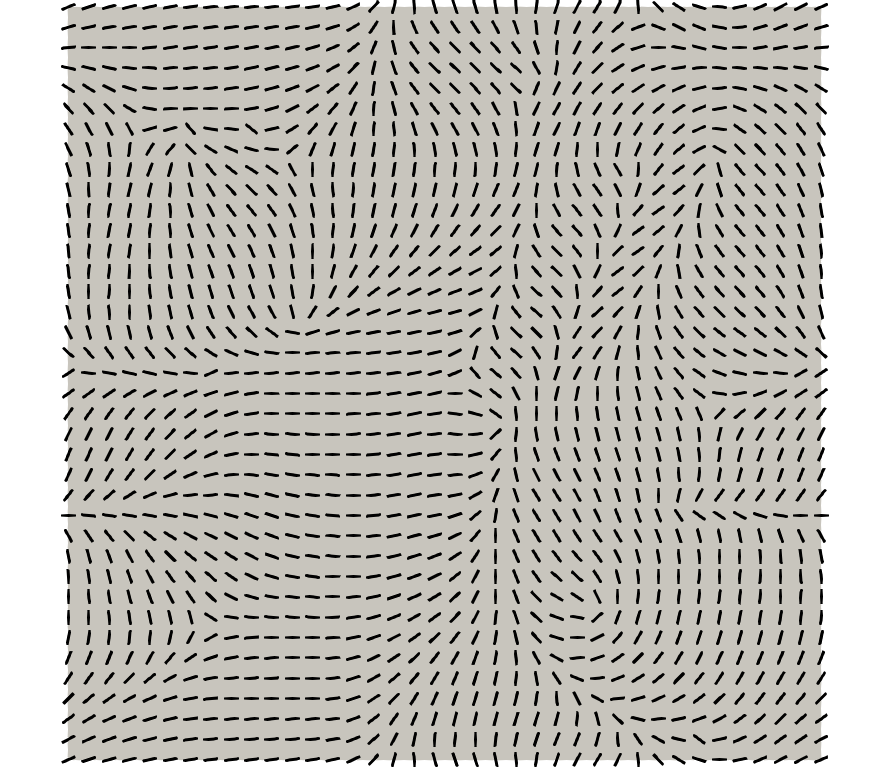}
\end{minipage}
\begin{minipage}{0.36\textwidth}
    \includegraphics[width=\textwidth]{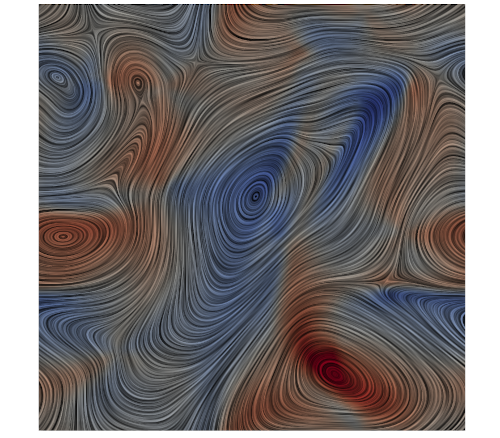}
\end{minipage}
\begin{minipage}{0.07\textwidth}
    \insertColorbarVertical{$\omega$}{$-6.5$}{}{}{$6.5$}
\end{minipage}
\caption{Active nematodynamics in square domain: (left) Q-tensor field $\mathbf{Q}$, visualized by the nomalized eigenvector of the largest eigenvalue. (right) Streamlines and vorticity $\omega = \operatorname{rot}\bulkVelocity$ of the velocity field $\bulkVelocity$. The used model parameters are $\lambda=0.1$, $S=1.133$, $\gamma=10$, $L=0.025$, $a=128.32$, $b=0$, $c=-200$, $\Re=1$ as well as $\alpha=5$. As in \cite{Pearceetal_PRL_2019} we have neglected the term including $\berisStress$ in the velocity equation due to the assumption that the active and viscous stresses dominate for the considered parameters.}
\label{fig1}
\end{figure}

The model is known from liquid crystal theory as an extension of the Beris-Edwards model \cite{Berisetal_Oxford_1994}, which is here supplemented with an additional active stress term and reads \cite{Giomi_PRX_2015}
\begin{align}
\dt\bulkVelocity + \left(\bulkVelocity\cdot\Grad\right)\bulkVelocity &= -\nablaTF\pressTF  + \frac{1}{\Re}\vecLaplace\bulkVelocity + \divTF\boldsymbol{\sigma} \\
\divTF\bulkVelocity &= 0\\
\dt\QTF + \left(\bulkVelocity\cdot\Grad\right)\QTF &= \lambda S\DTF + \OmegaTF \QTF - \QTF \OmegaTF + \gamma^{-1}\hTF
\end{align}
with fluid velocity $\bulkVelocity$, pressure $\pressTF$, Landau-de Gennes Q-tensor $\QTF$, order parameter $S$, flow alignment parameter $\lambda$, rotational viscosity $\gamma$, Reynolds number $\Re$, additional stress $\boldsymbol{\sigma} = \activeStress + \berisStress$ with active stress $\activeStress =  \alpha \QTF$ and an elastic stress $\berisStress = \QTF\hTF - \hTF\QTF - \lambda\hTF$, the molecular field $\hTF = L \vecLaplace \QTF - a \QTF + b ( \QTF^2 - \frac{1}{3} \trace (\QTF^2) \I) - c \trace (\QTF^2) \QTF$, deformation tensor $\DTF = \frac{1}{2} ( \nablaTF \bulkVelocity + (\nabla \bulkVelocity)^T)$ and vorticity tensor $\OmegaTF = \frac{1}{2} ( \nablaTF \bulkVelocity - (\nabla \bulkVelocity)^T)$, with $\alpha, L, a, b, c \in \R$. The model considers only a one-constant approximation of the Landau-de-Gennes energy (see \cite{Balletal_MCLC_2010} for more general expressions) and assumes only the simplest possible form of an active stress. We further have neglected the Ericksen stress. Supplemented with initial and boundary conditions the system without activity, $\alpha = 0$, has been analysed and numerically solved in \cite{Abelsetal_SIAMJMA_2014}. We here adapt this numerical approach and solve the system in a square domain in 2D with periodic boundary conditions. Fig. \ref{fig1} shows a time instant of a simulation.

\subsection{Multi-phase field model}
\label{ssec22}

For simplicity we consider a modeling approach without hydrodynamics. For hydrodynamic interactions we refer to \cite{Marthetal_IF_2016}. The model and numerical approaches to solve it have been introduced in \cite{Marthetal_JRSI_2015,Wenzeletal_JCP_2019}. Modified versions for active and passive systems can be found in \cite{Ziebertetal_RSI_2012,Camleyetal_PNAS_2014,Loeberetal_SR_2015,Camleyetal_JPD_2017,Muelleretal_PRL_2019} and \cite{Nonomura_PloSONE_2012}, respectively. We model each cell by a phase field active polar gel model \cite{Kruseetal_PRL_2004} and consider a short-range interaction potential between them. The evolution equations read
\begin{align}
\partial_t\phi_i + v_0\nabla\cdot\big(\phi_i\mathbf{P}_i\big) &= \gamma \Delta \mu_i\,, \\
\mu_i &= \frac{1}{Ca}\Big(\!-\epsilon\Delta\phi_i + \frac{1}{\epsilon}W'(\phi_i)\!\Big) + \frac{1}{Pa}\Big(\!\!-\frac{c}{2}\|\mathbf{P}_i\|^2 - \beta\nabla\cdot\mathbf{P}_i\!\Big) \nonumber\\
     & \quad + \frac{1}{In}\Big(\!B'(\phi_i)\sum_{j\neq i}w(d_j) + w'(d_i) d_i^\prime(\phi_i) \sum_{j\neq i}B(\phi_j)\!\Big),\\
\partial_t\mathbf{P}_i + \big(v_0\mathbf{P}_i\cdot\nabla\big)\mathbf{P}_i &= -\frac{1}{\kappa}\mathbf{H}_i\,, \\
\mathbf{H}_i &= \frac{1}{Pa}\Big(\!\!-c\phi_i\mathbf{P}_i + c\|\mathbf{P}_i\|^2\mathbf{P}_i - \Delta\mathbf{P}_i + \beta\nabla\phi_i\!\Big),
\end{align}
for $i = 1, \ldots, n$ phase field variables $\phi_i$ and polarization fields $\mathbf{P}_i$. The parameters $Ca$, $Pa$ and $In$ act as weightings between different energy contributions, a classical Ginzburg-Landau function with double-well potential $W(\phi) = \frac{1}{4}(\phi^2-1)^2$ and interface thickness $\epsilon$, a polar liquid crystal energy of Frank-Oseen type, with $c$ and $\beta$ parameters controlling the deformation of the polarization fields $\mathbf{P}_i$ and the anchoring on the cell interface, respectively, and the interaction term, which considers $B(\phi_i)=\frac{3}{\epsilon \sqrt{2}} W(\phi_i) \approx \delta_{\Gamma_i}$ as an approximation of the surface delta function for the cell boundary $\Gamma_i = \{ \mathbf{x}\in \Omega \;|\; \phi_i(\mathbf{x})= 0 \}$ and an approximation of an interaction potential $w(d_j)=\exp(- d_j^2 / \epsilon^2)$ with signed distance function $d_j(\phi_j) = -\frac{\epsilon}{\sqrt{2}}\ln((1+\phi_j(\mathbf{x}))/(1-\phi_j(\mathbf{x})))$ with respect to the zero-line (cell boundary $\Gamma_j$) of $\phi_j$. Activity is introduced in the evolution equations by a self-propulsion term, with velocity value $v_0$. For a detailed description of the numerical approach we refer to \cite{Praetoriusetal_NIC_2017}.

\begin{figure}[h]
\center
\includegraphics[width=0.32\textwidth]{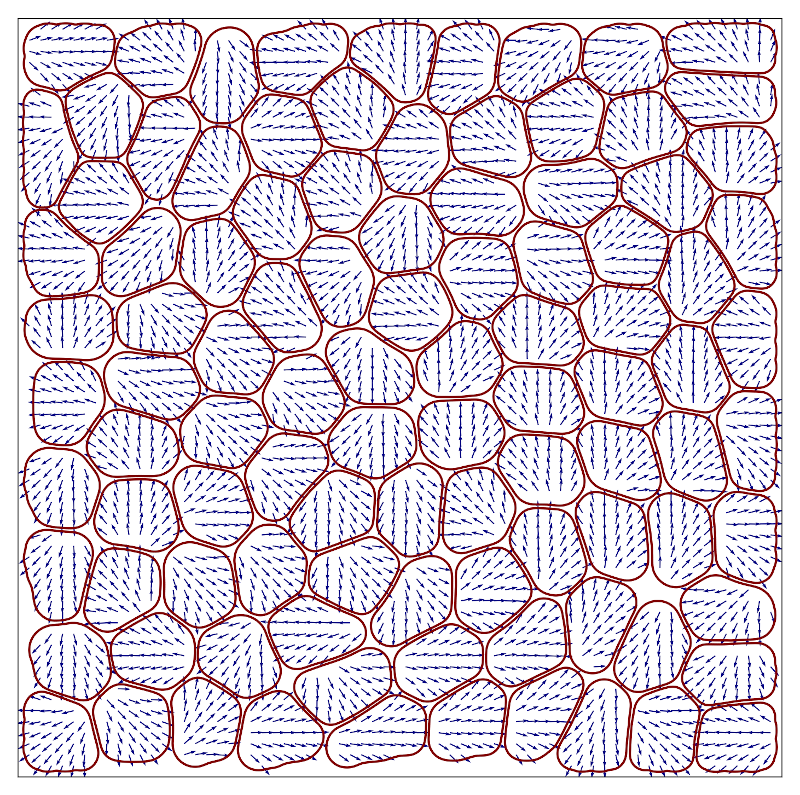}
\includegraphics[width=0.32\textwidth]{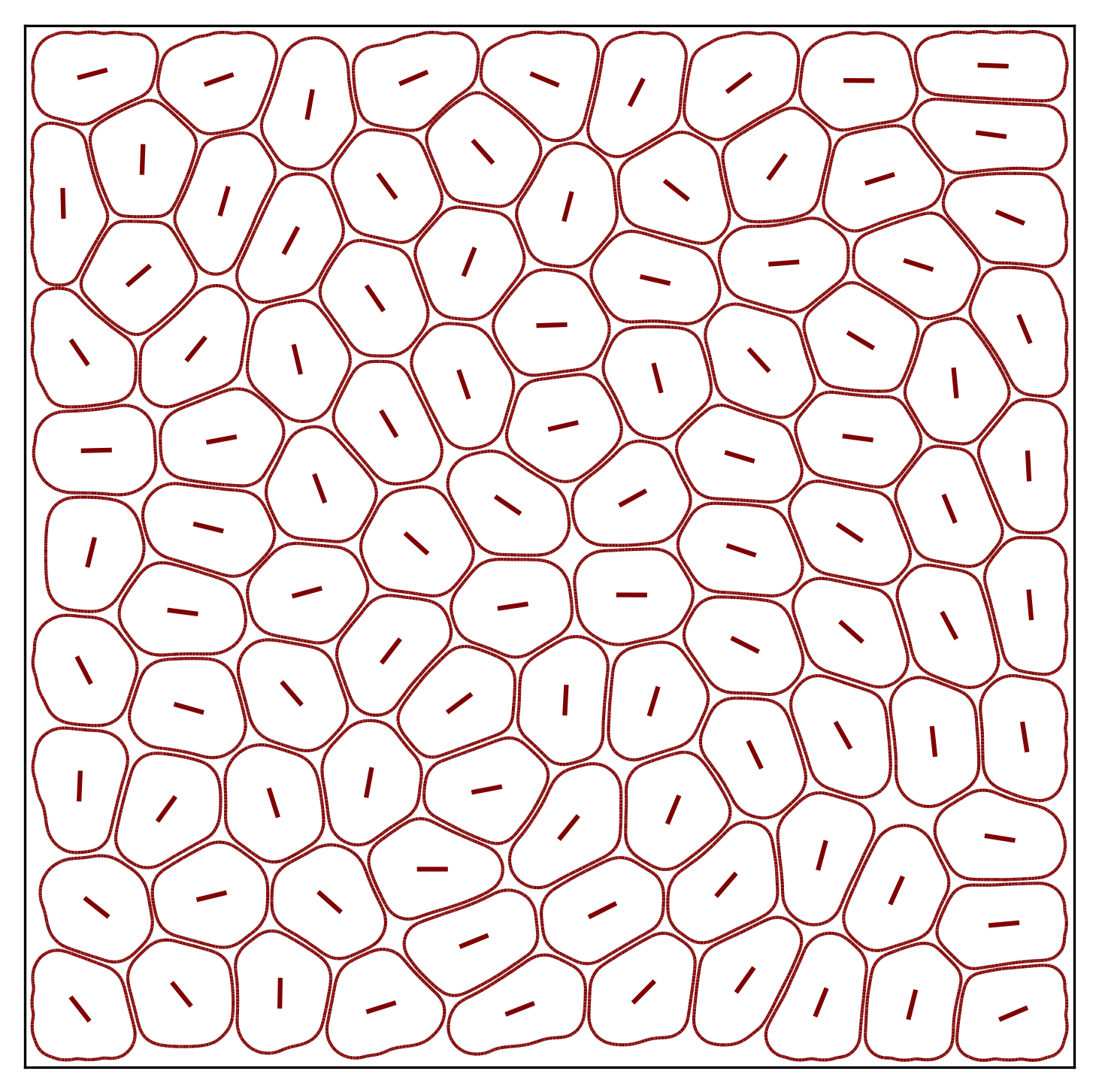}
\includegraphics[width=0.32\textwidth]{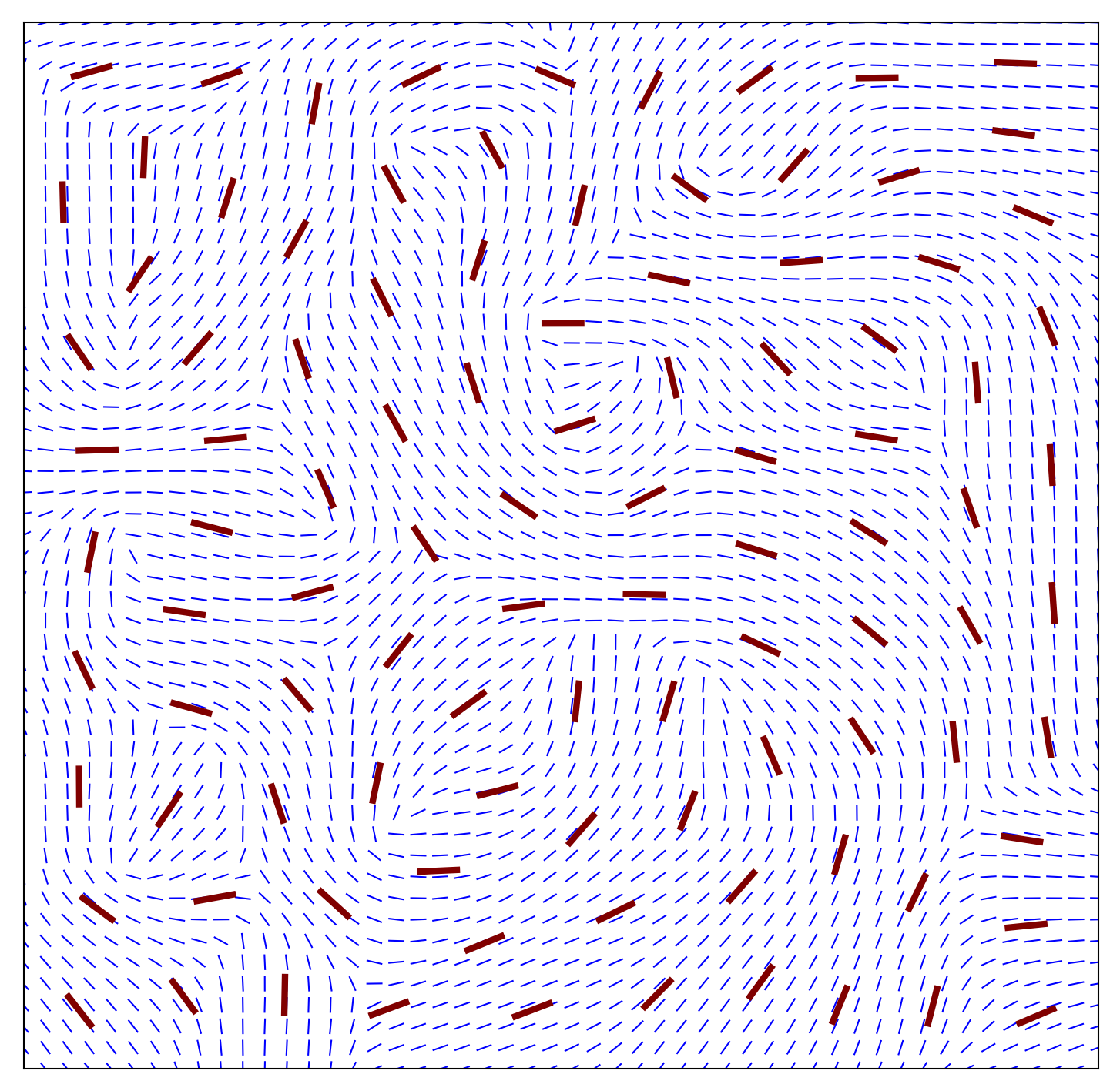}
\caption{Multi-phase field in square confinement: (left) Time frame with polarisation fields $\mathbf{P}_i$ and isocontours $\phi_i(\mathbf{x}) = 0$. (middle) Isocontours $\phi_i(\mathbf{x}) = 0$ and normalized eigenvector of $\mathbf{Q}_i$ corresponding to orientation of largest elongation in center of mass of each cell. (right) Interpolated Q-tensor field $\mathbf{Q}$, visualized by the normalized eigenvector corresponding to largest elongation with highlighted orientations in center of mass of each cell.}
\label{fig2}
\end{figure}

In order to analyse emerging nematic properties we define for each phase field variable $\phi_i$ a Q-tensor by
\begin{align*} 
		\mathbf{Q}_i = \begin{bmatrix}
		Q_{i,11} & Q_{i,12}\\
		Q_{i,12} & \!\!-Q_{i,11}
		\end{bmatrix} \!&=\! \int \!\begin{bmatrix}
		\frac{1}{2} \left((\partial_y \phi_i)^2-(\partial_x \phi_i)^2\right) & \!-(\partial_x \phi_i) (\partial_y \phi_i) \\
		-(\partial_x \phi_i) (\partial_y \phi_i) & \!\frac{1}{2} \left((\partial_x \phi_i)^2-(\partial_y \phi_i)^2\right)
		\end{bmatrix}d \mathbf{x}
\end{align*}
and obtain a continuous Q-tensor field $\mathbf{Q}$ by interpolating between these tensors in the center of mass in each cell. Fig. \ref{fig2} shows a time instant of a simulation with $n = 100$ cells. 

\section{Identification and tracking} 
\label{sec3}

\subsection{Topological defects}

\begin{figure}[h]
\center
  \begin{subfigure}[b]{.2\textwidth}
    \centering
    \begin{minipage}{\textwidth}
        \centering
        \includegraphics[width=.98\textwidth]{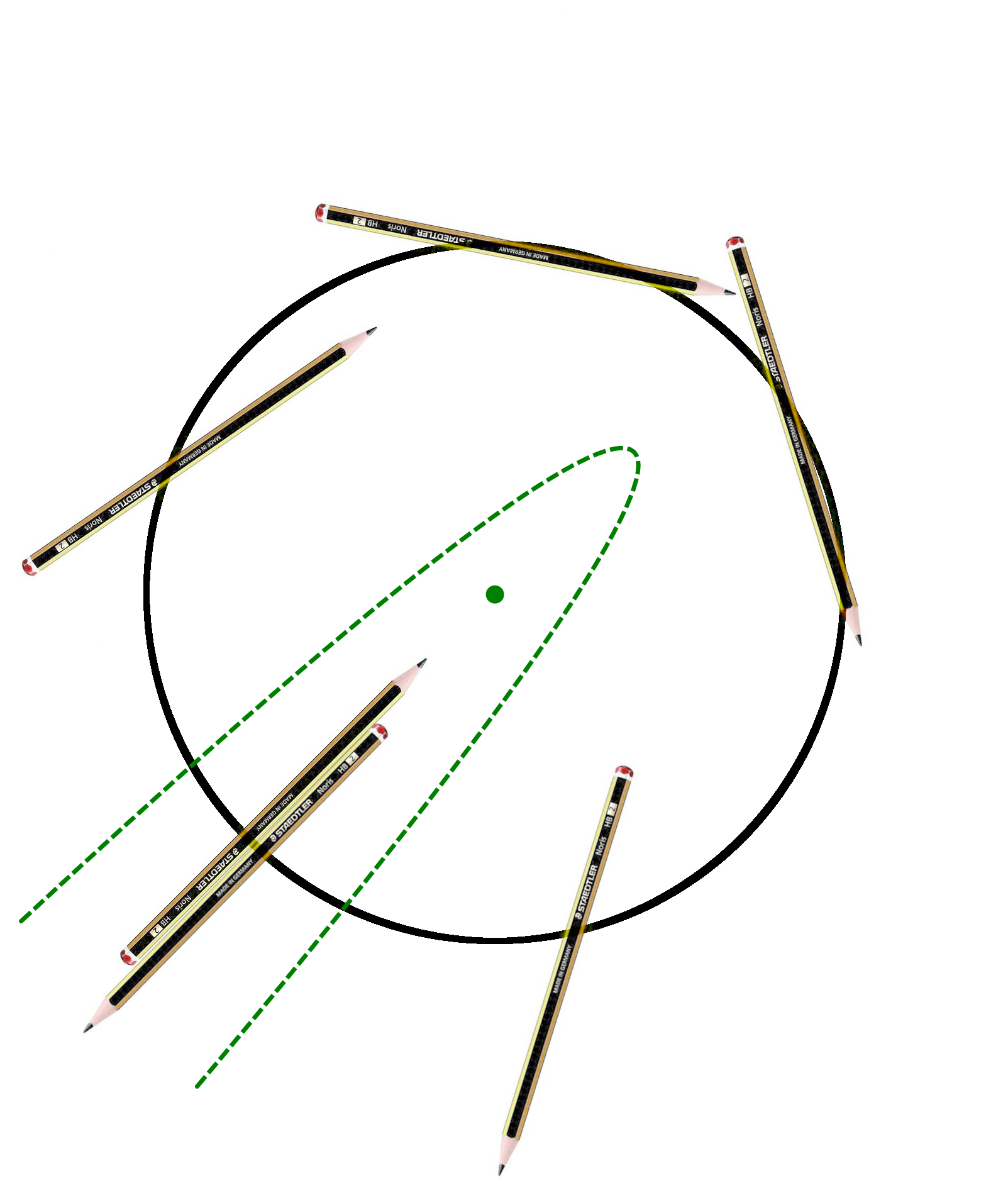}\\[4pt]
        \includegraphics[width=.98\textwidth]{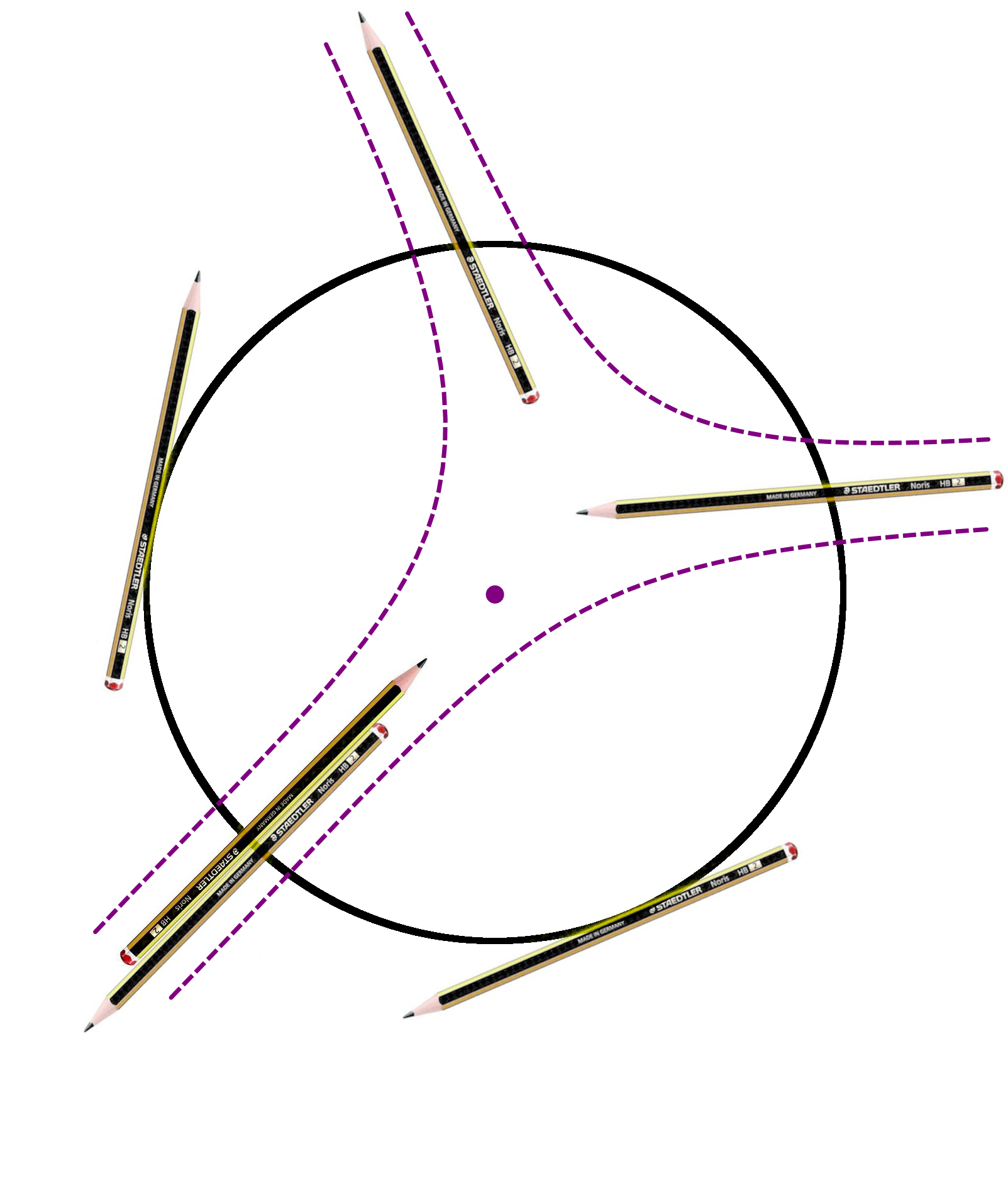}
    \end{minipage}
  \end{subfigure}
    \begin{subfigure}[t]{.32\textwidth}
    \centering
    \begin{minipage}{\textwidth}
        \centering
        \includegraphics[width=.98\textwidth]{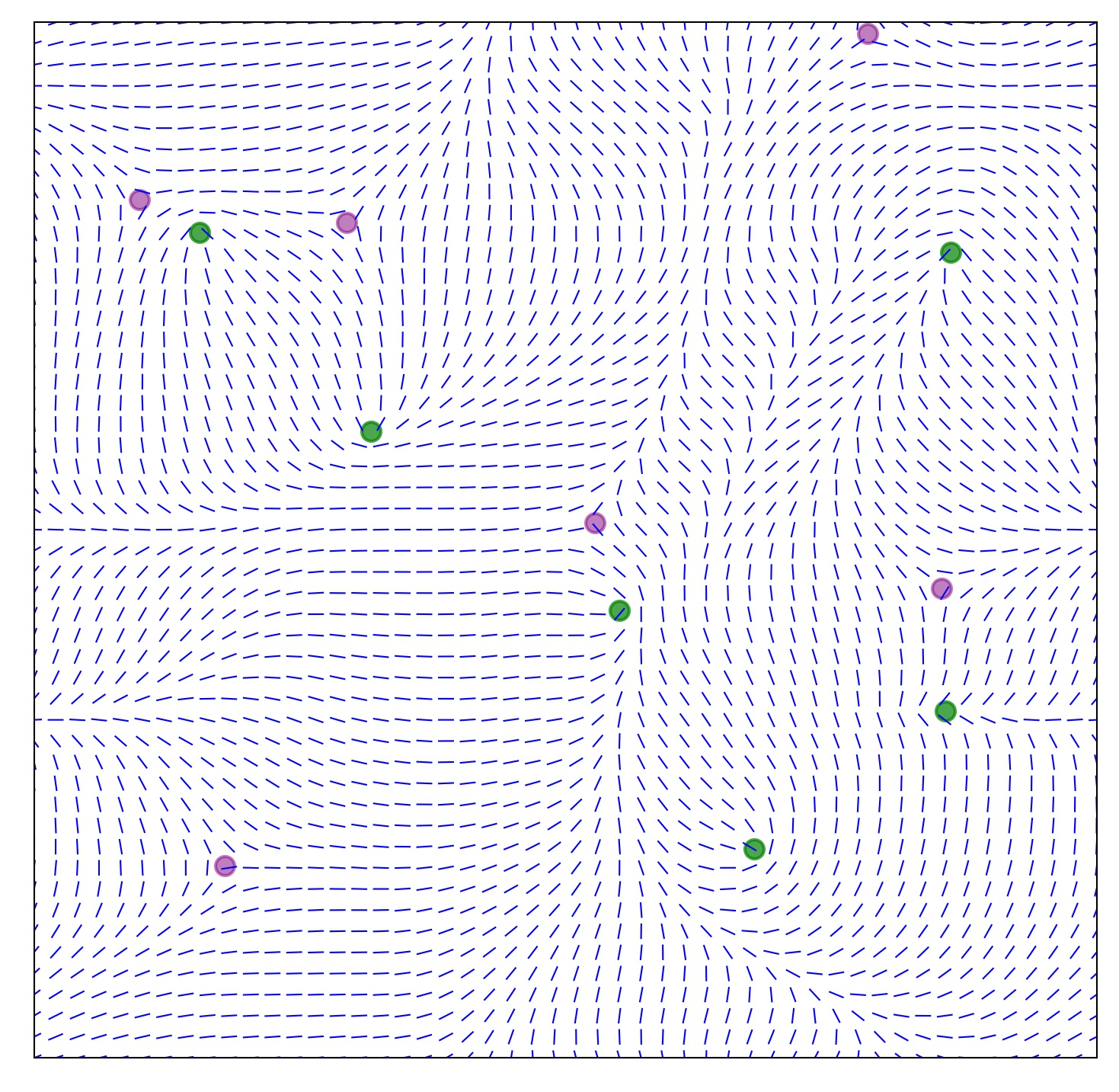}
    \end{minipage}
  \end{subfigure}
    \begin{subfigure}[t]{.32\textwidth}
    \centering
    \begin{minipage}{\textwidth}
        \centering
        \includegraphics[width=.98\textwidth]{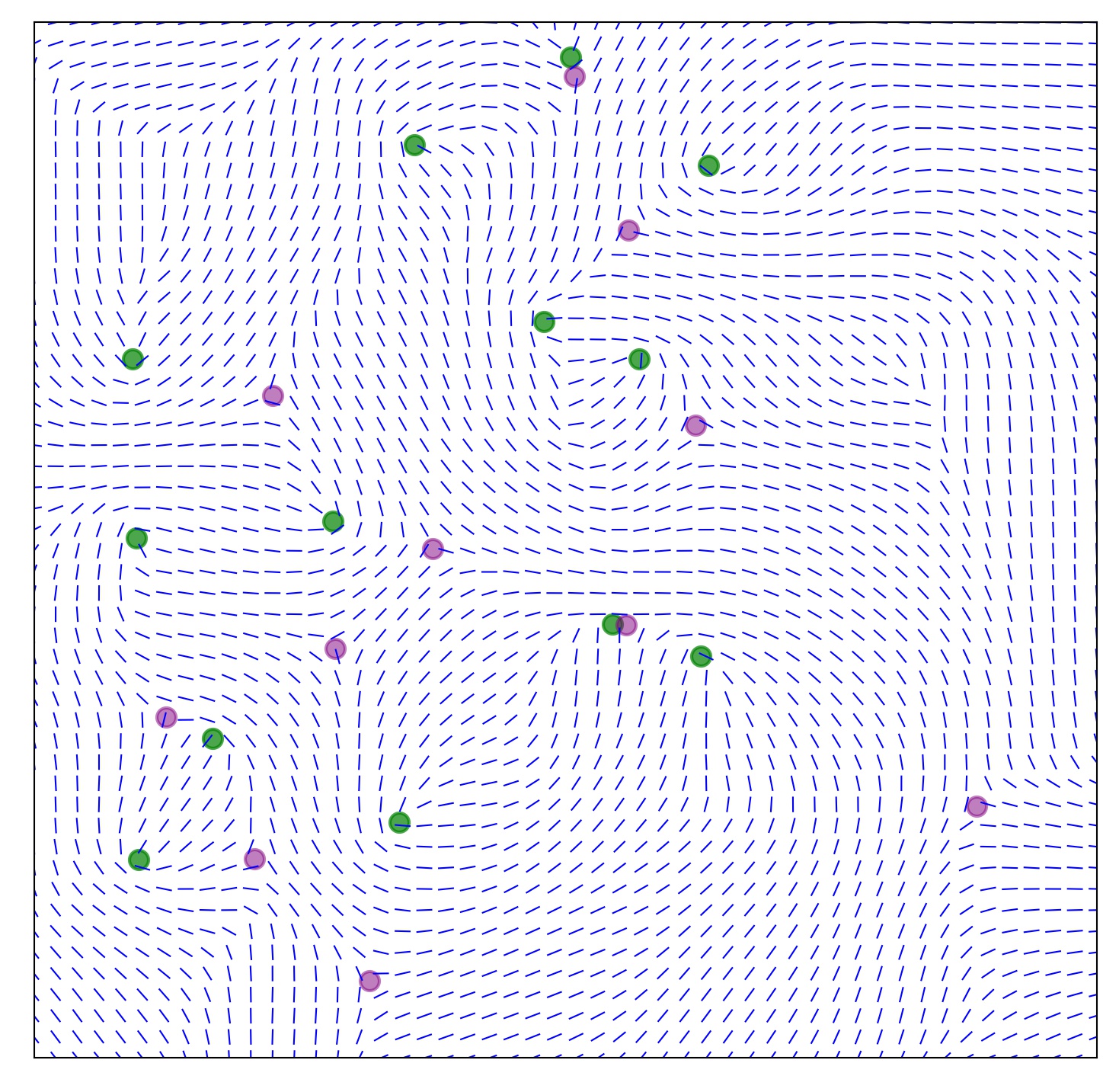}
    \end{minipage}
  \end{subfigure}
\caption{(left) Schematic description of $+ \frac{1}{2}$ and $- \frac{1}{2}$ defects. Locations for half-integer topological defects in the globalized director field from Fig. \ref{fig1} (middle) and Fig. \ref{fig2} (right), $+\frac{1}{2}$ (green), $-\frac{1}{2}$ (purple).}
\label{fig3}
\end{figure}

Topological defects are characterized as degenerated points of $\mathbf{Q}$ for which $Q_{11} = Q_{12} = 0$. For a nematic liquid crystal in 2D two types of topological defects predominate: comet-like ($+ \frac{1}{2}$) and trefoil-like ($- \frac{1}{2}$). The number associated with the topological defects, the winding number or topological charge, is the change in the orientation around the singular points along a full $2 \pi$ rotation, divided by $2 \pi$: for $+ \frac{1}{2}$ and $- \frac{1}{2}$ defects the orientation field rotates by $+ \pi$ and $- \pi$, respectively, see Fig. \ref{fig3}.

\subsection{Identification of defects} 

\begin{figure}[h]
  \centering
  \hspace{8pt}
  \begin{subfigure}[b]{.2\textwidth}
    \centering
    \begin{minipage}{\textwidth}
        \centering
        \includegraphics[width=.98\textwidth]{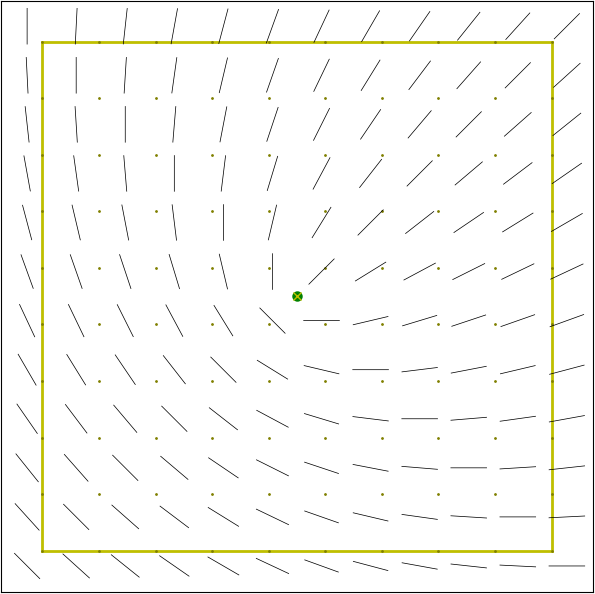}\\[4pt]
        \includegraphics[width=.98\textwidth]{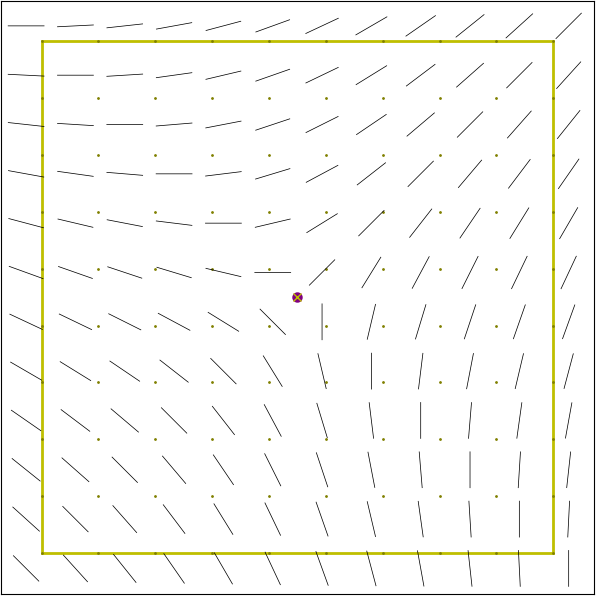}
    \end{minipage}
  \end{subfigure}
  \begin{subfigure}[t]{.7\textwidth}
    \centering
    \begin{minipage}{\textwidth}
        \centering
        \includegraphics[width=.98\textwidth]{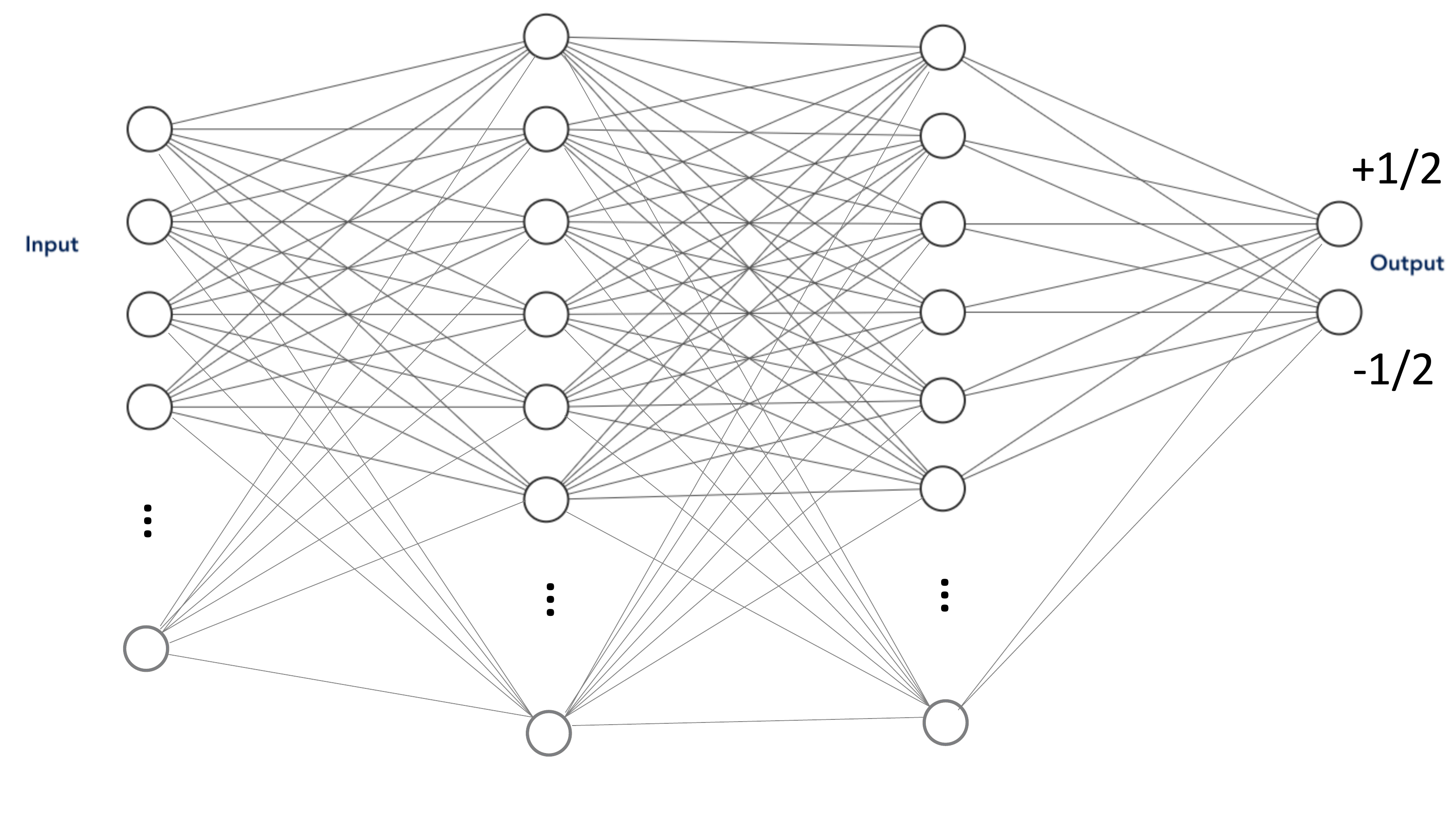}
    \end{minipage}
  \end{subfigure}
  \hspace{8pt}
  \caption{(left) Principle eigenvectors with defect in the center and kernel grid for interpolation for $+\frac{1}{2}$ (top) and $-\frac{1}{2}$ (bottom) defects. (right) ANN with kernel grid values as input (100) and defect type as output (2).}
  \label{fig:kernel_interpolation}
\end{figure}

We consider two approaches to identify the topological charge of the defects. The first is physics based and considers the sign of $\delta = \frac{\partial Q_{11}}{\partial x} \frac{\partial Q_{12}}{\partial y} - \frac{\partial Q_{11}}{\partial y} \frac{\partial Q_{12}}{\partial x}$ to distinguish between $+ \frac{1}{2}$ and $- \frac{1}{2}$ defects. The second treats the task as an image processing problem and uses an artificial neural network (ANN) to identify the defect type. The neural network consists of an input layer with 100 nodes, two hidden layers with 1000 and 200 nodes, respectively, and an output layer with two nodes, one for \(+\frac{1}{2}\) and \(-\frac{1}{2}\) defects. For every degenerated point of \(\mathbf{Q}\), a \(10\!\times\!10\) grid, called \textit{kernel}, with the defect located in the center, is computed by interpolation. This kernel data serves as input, see Fig.~\ref{fig:kernel_interpolation}. The training data is constructed analytically by specifying boundary conditions for vector fields. We consider 5000 simulated defects for each topological charge. To generate a decent variability, the defect is affected by surrounding defects and noise. Each data is trained within five epochs with a learning rate of \(0.05\). Tests on ideal, single affected and multiple combined defect data have been performed leading to identification rates of \(96.98\,\%\). More comprehensive network constructions and more extensive training processes would lead to higher precision. Applied to the simulation data, the classification with the ANN yields an identification rate of \(99.31\,\%\). 

\subsection{Tracking of defects} 

With known defect positions and type in each time frame we have to connect them from frame to frame. Dozens of software tools have been developed for this task in the context of particle tracking \cite{Meijeringetal_ME_2012}. For a comparison of these methods we refer to \cite{Chenouardetal_NM_2014}. We here use an approach described in \cite{Sbalzarinietal_JSB_2005}. It involves finding a set of associations between the defect locations in subsequent frames such that a cost functional is minimized. It is based on a particle matching algorithm using a graph theory technique. It allows to consider different defect types and defect appearance and disappearance and is available as a plugin for ImageJ and Fiji (www.imagej.net) \cite{Schindelinetal_NM_2012}.

\section{Results}
\label{sec4}

\subsection{Simulation data}

\begin{figure}[h]
  \centering
  \hspace{8pt}
  \begin{subfigure}[t]{\textwidth}
    \centering
    \includegraphics[width=.7\textwidth]{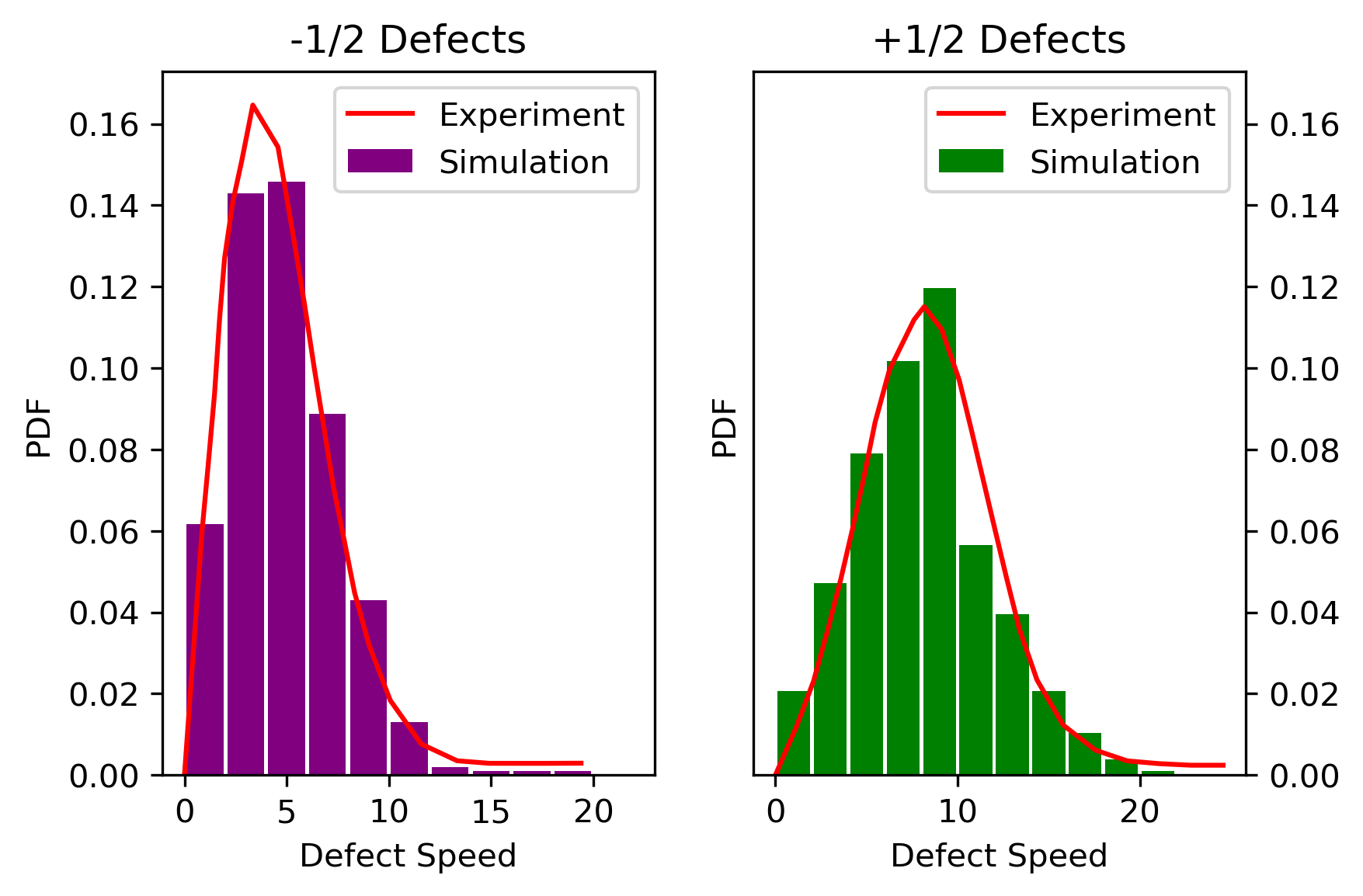}
  \end{subfigure}
  \begin{subfigure}[t]{\textwidth}
    \centering
    \includegraphics[width=.7\textwidth]{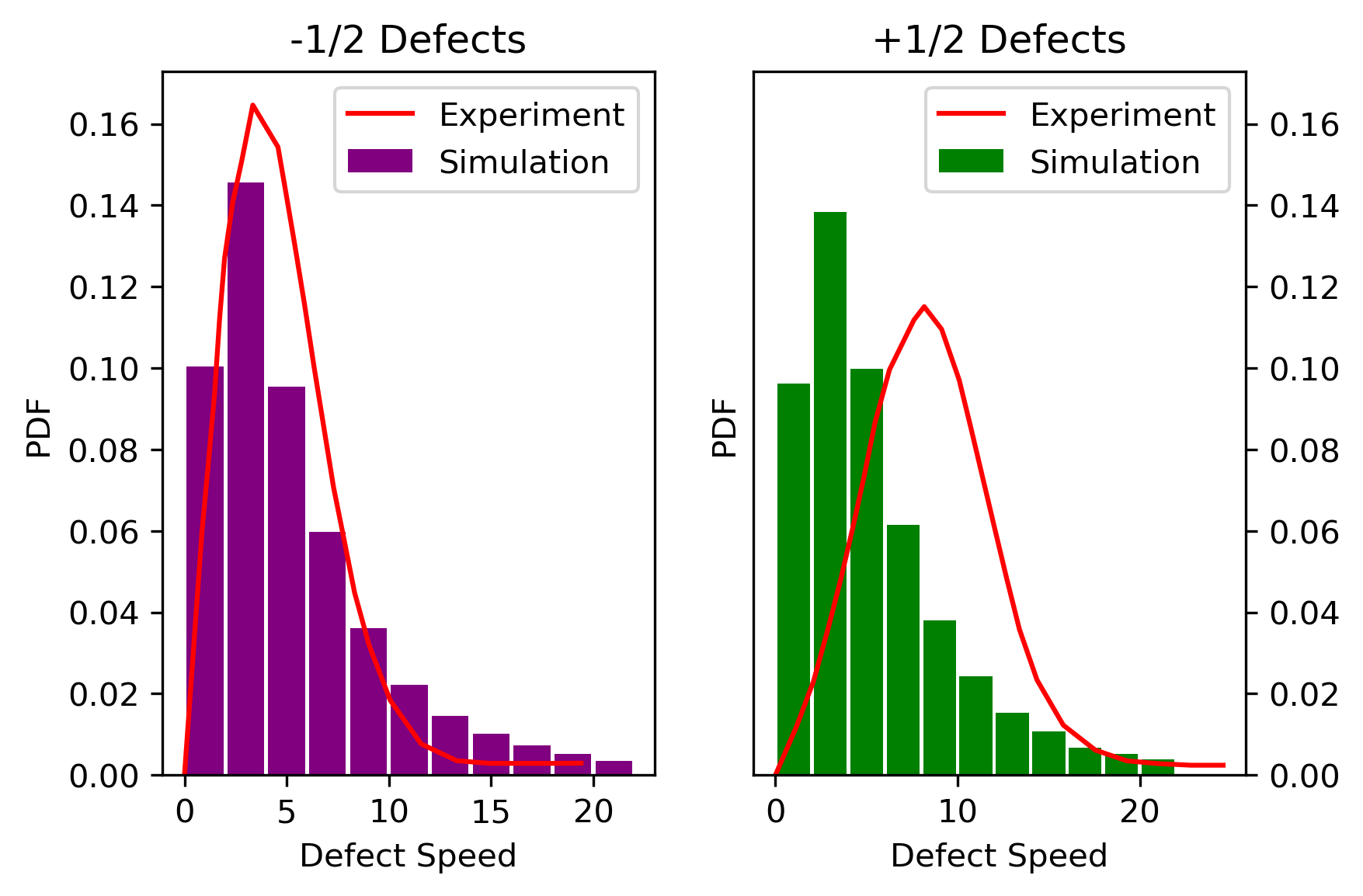}
  \end{subfigure}
  \hspace{8pt}
  \caption{Velocity distribution of topological defects in the active nematodynamic (top) and multi-phase field (bottom) model, both in comparison with experimental data from \cite{decamp_redner_baskaran_hagan_dogic_2015} for $+ \frac{1}{2}$ and $- \frac{1}{2}$ defects.}
  \label{fig:histograms}
\end{figure}

Using the described tools for defect identification and tracking, we can statistically examine the velocity distribution of topological defects. We compute these data for both models and compare them with experimental data. However, due to a lack of available data for defects in epithelia tissue, we here compare both with data for MT bundles \cite{decamp_redner_baskaran_hagan_dogic_2015}. Similar comparisons, with a higher order Landau-de Gennes model have been performed in \cite{Oza_2016}. As both of our models are written in dimensionless units we first rescale our data based on the reported average velocity of $6.6 \mu m s^{-1}$ in \cite{Oza_2016}. Figure \ref{fig:histograms} shows the comparison for both models for $+ \frac{1}{2}$ and $- \frac{1}{2}$ defects. We observe an excellent agreement of the experimental data with the active nematodynamics model of section \ref{ssec21}. The results also show a significant difference in the velocity distribution between $+ \frac{1}{2}$ and $- \frac{1}{2}$ defects. $+ \frac{1}{2}$ defects are significantly faster, which again is in agreement with experimental data and theoretical predictions \cite{Doostmohammadietal_NC_2018}. This difference is still present in the multi-phase field model of section \ref{ssec22}, but much less pronounced. The velocity distributions for $+ \frac{1}{2}$ and $- \frac{1}{2}$ defects are almost equal to each other in this model. 
 
We can also analyse the direction of defect movement. In the active nematodynamics model $+ \frac{1}{2}$ defects move in the direction of the 'tail' of the defect, indicating contractile systems. The direction of movement can be tuned by the sign of $\alpha$ in the active stress. For the multi-phase field model, this is less obvious, as the defects are a secondary effect and not directly related to movement. However, also in these systems contractile behavior is most common, which again is in agreement with experimental measurements \cite{Sawetal_N_2017}.  

\begin{figure}[h]
  \centering
  \hspace{8pt}
  \begin{subfigure}[t]{.45\textwidth}
    \centering
    \includegraphics[width=.8\textwidth]{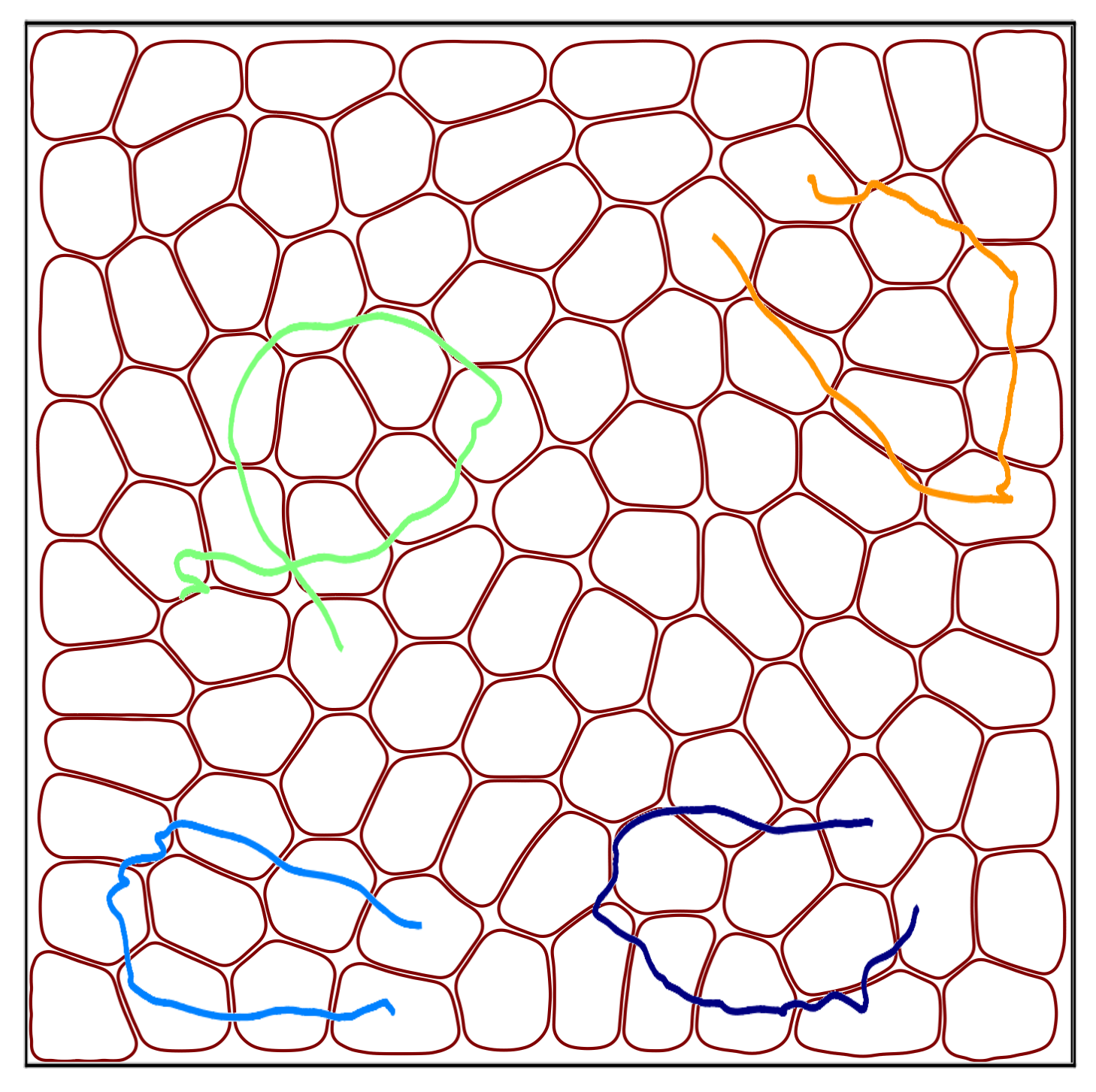}
  \end{subfigure}
  \begin{subfigure}[t]{.45\textwidth}
    \centering
    \includegraphics[width=.8\textwidth]{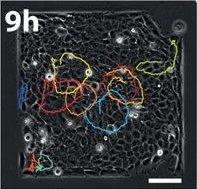}
  \end{subfigure}
  \hspace{8pt}
  \caption{Oscillations: (left) Isocontours $\phi_i(\mathbf{x}) = 0$ for the multi-phase field model. (right) Microscopy images of HuCaT cells in square confinement, taken from \cite{Peyretetal_BJ_2019}(Fig. 1). Both images additionally show the trajectories of the centers of mass for certain cells over some time period, highlighting the sustained oscillations of cells.}
  \label{fig:oscillations}
\end{figure}

Another example which shows qualitative agreement with experimental data is achieved by considering the multi-phase field model described in Section \ref{ssec22} with ~100 cells in a square confinement. Within a certain activity range this leads to oscillations in the cell movements. These oscillations are in qualitative agreement with results on microscopy images of human keratinocytes (HaCaT cells) in similar confinements, see \cite{Peyretetal_BJ_2019}. Figure \ref{fig:oscillations} visualizes the trajectories of the centers of mass for some cells and compares them with the experimental data. 
\\

\subsection{Experimental data}

The same methodology to identify, classify and track defects is applicable to experimental data. Instead of the numerical solution it just requires an image processing step to map microscopy images to Q-tensor fields. For epithelia tissue this requires a classical segmentation problem to identify each cell, which than can be represented by a phase field variable, which can be processed as in Sec. \ref{sec2}. For filamentous particles, e.g. MT bundles the directional field has to be estimated. This can be done be computing a gradient vector on the gray-scale image for each pixel and appropriately averaging by first doubling the angle and squaring the length to account for the head-tail symmetry. In \cite{Bazanetal_IEEE_2002} it is shown that this approach is equivalent to a principal component analysis and robustly leads to directional fields $\mathbf{n}$. The Q-tensor field can than be computed by $\mathbf{Q} = S (\mathbf{n} \otimes \mathbf{n} - \frac{1}{2} \mathbf{I})$ with $S$ the scalar order parameter, encoding the degree of alignment with the average direction and $\mathbf{I}$ the identity matrix. Detailed comparisons on defects between experimental and simulation data will be done elsewhere based on the proposed algorithms.

\section{Conclusion}
\label{sec5}

We here provide reliable and robust data analysis tools to identify, classify and track topological defects in simulation and microscopy data. We thereby concentrate on $\pm \frac{1}{2}$ defects in active nematics. However, the tools can easily be adapted to other defects, e.g. $\pm 1$ defects in polar system. We consider both, physics based approaches as well as ANN, to classify the defects. Especially if applied to microscopy data, we see advantages of the ANN approach to deal with noise. Here we only consider simulation data for two prototypical models, an active nematodynamics model, e.g. applicable to model MT bundels, and a multi-phase field model to simulate epithelia cell sheets. Qualitative properties and statistical data on defect velocities are computed and compared with experimental results from the literature. Detailed quantitative comparisons will be done elsewhere.     

\begin{acknowledgement}
AV acknowledges financial support from DFG through FOR3013. The work was also supported by the Sino-German Science Center on the occasion of the Chinese-German Workshop on Computational and Applied Mathematics in Kiel 2019. We further acknowledge computing resources provided by JSC under grant HDR06 and ZIH/TU Dresden.
\end{acknowledgement}

\bibliographystyle{unsrt}
\bibliography{lit}
\end{document}